\documentclass[twocolumn,elsart1p.cls,floatfix]{revtex4}
\usepackage{color}
\usepackage{epsfig}

\begin{document}

\title{ Melting of alloys along  grain boundaries}
\author{Efim A. Brener and D. E. Temkin}
\affiliation{Institut f\"ur Festk\"orperforschung, Forschungszentrum J\"ulich,
D-52425 J\"ulich, Germany} 

\begin{abstract}
We discuss melting of alloys  along  grain boundaries as a free boundary problem 
for two moving solid-liquid interfaces. One of them is the melting front and the other 
is the solidification front. 
The presence of the triple junction plays an important role in controlling 
the velocity of this process.
The interfaces  strongly interact 
via the diffusion field in the thin liquid layer between them. 
In the liquid film migration (LFM) mechanism the system chooses 
a  more efficient kinetic path, which is controlled by  diffusion in the liquid film 
on   relatively short distances. However,
only  weak coherency strain energy is the effective driving force for LFM 
in the case of  melting of one-phase alloys. 
 The  process with only one melting front would be controlled
 by the very slow diffusion in the mother solid phase on   relatively large distances.

\end{abstract}

\maketitle

\section{Introduction}
In our previous paper \cite{brener06prl} we discuss  partial melting 
 of the binary alloys which proceeds via the liquid film migration (LFM) mechanism.
The liquid phase separates  two nearly parabolic fronts  (see Fig. 1).
If only the melting front existed,  
the  process would be controlled by the very slow diffusion 
in the mother solid phase. 
In the LFM mechanism the system chooses 
a  more efficient kinetic path  which 
 is controlled by  much faster diffusion in the liquid film. However, 
in this case the relatively weak coherency strain energy is  involved as  
an effective driving force for this process. 
\begin{figure}
\begin{center}
\epsfig{file=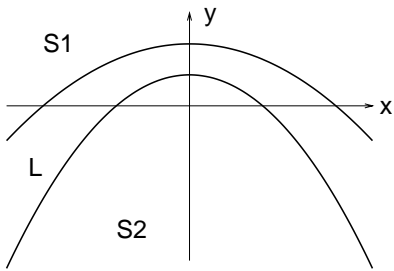, width=6cm}
\caption{Schematic representation of two  moving nearly parabolic fronts 
discussed in Ref. \cite{brener06prl}; $S1$ and $S2$ are the melting and growing 
solids, and $L$ is the liquid film.}
\end{center}
\end{figure}

The early observations of liquid film migration (LFM) were made
during sintering in the presence  of liquid phase \cite{yoon1} 
or during partial melting of alloys \cite{musch1}  
(see  \cite{yoon2} for a review).  Nowadays LFM
is a well established phenomenon of great practical importance.  
In LFM one  crystal is melting
and another one is solidifying.  Both solid-liquid interfaces 
move  together with the same velocity. 
The migration 
velocity is much smaller than the characteristic velocity of 
atomic kinetics at the interfaces. Therefore,  both solids should at the 
interfaces  be  locally in thermodynamic equilibrium with the  
liquid phase. On the other hand,  these local equilibrium 
states should be different for the two interfaces 
to provide the driving force for the process. 
 It is by now well accepted  (see, for example, \cite{yoon2,4}) 
that the difference of the equilibrium states at the melting and solidification fronts 
is due to the coherency strain energy,  
important only at the melting front because of the sharp concentration profile  
ahead the moving melting front (diffusion in the solid phase is very slow and the 
corresponding diffusion length is very small). The solute atoms diffuse ahead 
of the moving film and the coherency strain energy in such frontal diffusion zone 
arises from the solute misfit.
Thus, the equilibrium liquid composition
at the melting front, which depends on the coherency strain energy 
and on the curvature of the front, differs from the liquid composition 
at the unstressed and curved solidification front. This leads to the 
necessary gradient of the concentration across the liquid film and the process is 
controlled by the diffusion in the film.


Thus, a theoretical description of LFM requires the solution
of a free boundary problem for two combined moving solid-liquid 
interfaces with a liquid film in between.
In Ref.\cite{temkin2005}  
this problem was considered for simplified boundary conditions: 
the temperature and the chemical composition along each interface 
were kept constant. 
This means that any capillary, kinetic and crystallographic 
effects at the interfaces were neglected. 
It was found that under these simplified 
boundary conditions two co-focal parabolic fronts can move together 
with the same velocity. The situation is rather similar to a steady-state  motion 
of one parabolic solidification front into a supercooled melt 
or one parabolic melting front into a superheated solid. 
In this approximation the Peclet numbers were found, but the steady-state velocity 
remained undetermined at that stage. Thus, the problem of velocity selection arises. 

Solvability theory has been very successful in predicting certain properties of 
 pattern selecting in dendritic growth and a number of related phenomena 
(see, for example, \cite{saito96,kessler88,brener91}).
We note that capillarity is 
a singular perturbation and the anisotropy of the surface energy is a prerequisite 
for the existence of the solution in this theory. 
In \cite{brener06prl} we extended the selection theory for the process of liquid film 
migration where the strong diffusion interaction between melting and solidification 
fronts plays a crucial role. 

The nucleation of the melt often takes place at the grain 
boundaries of the mother solid phase. Then the thin liquid layer extends along the 
grain boundary. 
In the present paper we discuss the liquid film migration during the partial melting 
along the grain boundary (see Fig.2). The presence of the triple junction in this 
geometry drastically changes the structure of the theory. It produces a very strong 
perturbation of the solid-liquid interface and the anisotropy of the surface tension 
does not play an important role in such processes 
\cite{temthes,brener07prl,brener07acta}.
 
\section{Formulation of the problem}

\begin{figure}
\begin{center}
\epsfig{file=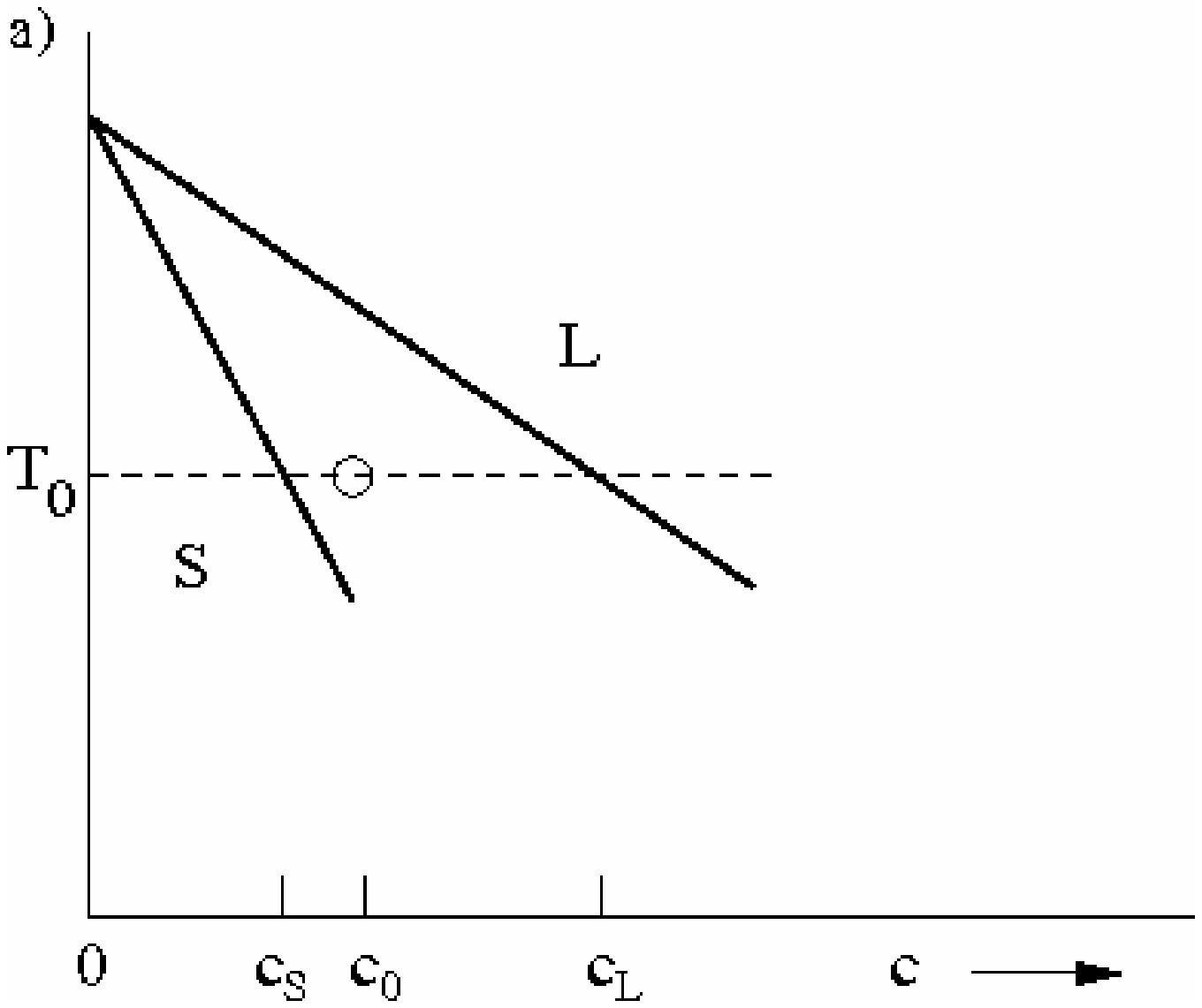, width=6cm}
\epsfig{file=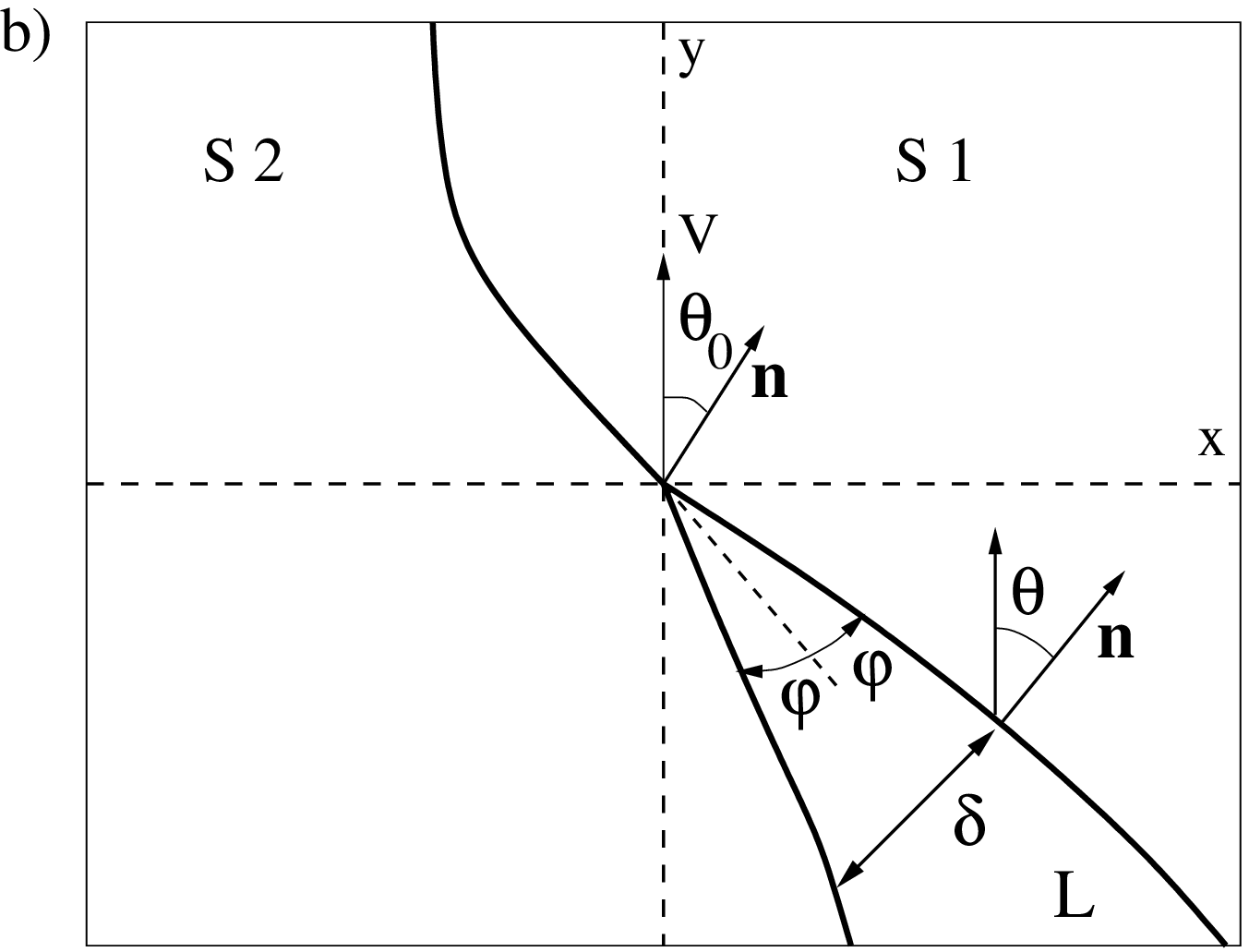, width=6cm}
\caption{Schematic representation of the phase diagram a) and 
and configuration of different interfaces near the triple junction b).
This structure moves steadily with the velocity $V$ in the vertical $y$ direction. 
The triple junction is located at the origin. Above the origin two solid grains 
$S1$ and $S2$ are separated by the grain boundary. Below the origin two grains 
are separated by the thin liquid layer $L$.}
\end{center}
\end{figure}

We discuss  the two-dimensional problem of the steady-state motion of a 
thin liquid film during the process of 
isothermal melting of a binary alloy along the grain boundary. The schematic phase 
diagram of the binary alloy and the geometry of the process are presented in Fig.2.
We note that the discussed geometry corresponds to  the spontaneous breaking 
of the reflection symmetry $x\rightarrow -x$. 
The equivalent structure can be obtained by the 
reflection $x\rightarrow -x$ of the structure presented in Fig.2b. 
We assume that the diffusion in 
the solid phases is very slow and the concentration $c$ in the liquid film obeys the 
Laplace equation. We introduce the normalized concentration $C=(c-c_L)/(c_L-c_S)$
with $c_L$ and $c_S$ being the  liquidus and solidus 
concentrations of the equilibrium phase diagram at a given temperature, $T_0$.
Then the equilibrium concentration and the mass balance conditions
at the solidification front, which separates the grain $S2$ and the liquid phase, read
\begin{equation}
C_2=-d_0K_2, \qquad  V_n=-D\partial C/\partial \bf{n}.
\label{solid}
\end{equation}
At the melting front, which separates the grain $S1$ and the liquid phase, 
the equilibrium concentration is changed 
by the presence of the elastic coherency strain energy \cite{yoon2} 
and  also the diffusional flux changes 
 because in the solid ahead of the melting front 
the concentration is $c_0$ which is different from $c_S$:
\begin{equation}
C_1=-b\Delta^2+d_0K_1, \qquad  V_n(1-\Delta)=-D\partial C/\partial \bf{n}.
\label{melt}
\end{equation}
Here $V_n$ is the normal velocity; 
$D$ is the diffusion coefficient in the liquid film; $K$ is the curvature assumed 
to be positive for the interfaces in Fig.1; $\Delta=(c_0-c_S)/(c_L-c_S)$ is the 
dimensionless driving force; $b=Y\Omega(da/dc)^2/a^2f_L''$ 
is the dimensionless constant which describes the 
coherency strain energy \cite{4}, $\Omega$ is the atomic volume, 
$Y$ is the bulk elastic modulus,  $a$ is the atomic constant,
$f_L(c)$ is the free energy of the liquid phase per atom,
$f_L''$ is the second derivative of $f_L(c)$ at $c=c_L$; 
$d_0$ is the  chemical capillary length which is assumed to be isotropic in the 
present problem,
$d_0=\gamma\Omega / f_L''(c_L-c_S)^2$
where $\gamma$ is the surface energy of the solid-liquid interface.

At an arbitrary rotation of  the grain boundary
 (Fig. 2) the angle between fronts at 
the triple junction $2\varphi$ remains unchanged, $\cos\varphi=\gamma_b/2\gamma$, 
where $\gamma_b$ is the surface tension of the grain boundary. 
This additional thermodynamical condition eventually allows to find a unique solution 
for the whole structure and the velocity of steady-state propagation of this structure.
 
In the general case, the 
formulated problem is quite complicated because two fronts interact non-locally 
through the diffusion field inside of the liquid phase.  
At small driving forces $\Delta$ and small angles $2\varphi$ between fronts 
this problem can be treated in the so-called ``lubrication'' approximation 
which is also called the boundary layer model (BLM) in the context of solidification. 
This allows to reduce the original nonlocal problem to the system of ordinary 
differential equations for the shapes of the  fronts. 

\section{ Lubrication approximation}
The  crucial ingredient of the lubrication  approximation is that the variation of the 
concentration field in the thin liquid layer
in the direction normal to the interfaces  is much larger than in the tangential 
direction. 
In this approximation the two fronts are close to each other and they can be parametrized
by two functions which are invariant to the  transformation of the coordinate system:
$K(\theta)$ and $\delta(\theta)$. $K$ is the curvature 
of  one of the interfaces and $\delta$ is the distance 
between the interfaces in the normal direction;  
$\theta$ is the angle between the normal to  one of interfaces $\bf{n}$
and the  direction of the steady-state growth (see Fig.2)
Alternatively, one can use functions 
$\theta(s)$ and $\delta(s)$ where $s$ is the arc-length along the interface 
and $K=\partial {\theta}/\partial {s}$. The lubrication approximation is valid if 
$$K\delta\ll 1$$ which is the small parameter of the theory. 
We will see later that this condition is fulfilled if $\Delta$ and $\varphi$ are 
small.

Let us derive the equations for the solidification and melting fronts using  basic 
Eqs. (\ref{solid}, \ref{melt}) and the crucial ingredient of the lubrication 
approximation that the concentration field varies linearly in the normal direction. 
Then in the main approximation $\partial{C}/\partial{\bf n}\approx (C_1-C_2)/\delta$ 
and from the mass balance condition, Eq. (\ref{solid})  we find 
\begin{equation}
V\cos\theta=D\frac{b\Delta^2-2d_0K}{\delta},
\label{V}
\end{equation}
where $V$ is the steady-state velocity and we have already used the fact 
that the interfaces are close to each other: $K_1\approx K_2=K$.
The derivation of the second equation is slightly more delicate, because if one takes 
the difference between two mass balance conditions all remaining terms are small 
and one should take into account the small differences between normal velocities 
and normal gradients at two interfaces:
\begin{equation}
V_{n1}-V_{n2}+D\left[\left(\frac{\partial{C}}{\partial{\bf n}}\right)_1- 
 \left(\frac{\partial{C}}{\partial{\bf n}}\right)_2\right]=V\Delta\cos\theta. 
\label{differ}
\end{equation}
Evaluating the term in the right-hand-side which is already proportional to the small 
parameter $\Delta$ we have ignored the difference between normal velocities at two 
interfaces. 
In the left-hand-side, the small difference in normal velocities is due to the 
differences in the normal directions at the two interfaces: 
\begin{equation}
(V_{n1}-V_{n2})\approx V\sin\theta\partial{\delta}/\partial{s}
=VK\sin\theta\partial{\delta}/\partial{\theta}.
\label{Vdif}
\end{equation}
The small difference in normal gradients at the two interfaces is due to the curvature 
of the interface. The simple way to see  this effect is to look for the 
solution of the Laplace equation 
for the concentration field $C=A+B\ln r $  in the local polar coordinate system  
rather than in the 
Cartesian coordinates as it would be convenient for  flat interfaces.   
One of the interfaces is located at $r=1/K(\theta)$ and the other is at 
$r=(1+K\delta)/K$.
Then using the fact that the gradient scales as $1/r$ and
 $K\delta\ll 1$ we obtain:
\begin{equation}
D\left[\left(\frac{\partial{C}}{\partial{\bf n}}\right)_1- 
 \left(\frac{\partial{C}}{\partial{\bf n}}\right)_2\right]\approx 
-D\left(\frac{\partial{C}}{\partial{\bf n}}\right) K\delta=VK\delta\cos\theta.
\label{graddif}
\end{equation}
 
Inserting Eqs. (\ref{Vdif}, \ref{graddif}) into Eq. (\ref{differ}) we obtain:
\begin{equation}
\frac{\partial}{\partial{s}}\left[\delta\sin\theta\right]=
K\frac{\partial}{\partial{\theta}}\left[\delta\sin\theta\right]
=\Delta\cos\theta. 
\label{delta}
\end{equation}

Eqs. (\ref{V}) and  (\ref{delta}) are the desired equations in the leading order 
of the lubrication approximation. As seen from Eq.(\ref{delta}) 
$K\delta\sim\Delta\ll1$ which justifies the  approximation used in the 
limit of small driving forces.  

At the triple junction, where $\delta=0$ and $\partial{\delta}/\partial{s}=2\varphi$,
Eq. (\ref{delta}) reads:
\begin{equation}
\tan\theta_0=\frac{\Delta}{2\varphi},
\label{triple}
\end{equation}
where $\theta_0$ is the value of $\theta$ at the triple junction. The grain boundary 
is rotated by approximately the same angle $\theta_0$ at the triple junction 
(see Fig. 2). 
One can eliminate the curvature $K$ from Eqs. (\ref{V}, \ref{delta}) and obtain 
the closed first order differential equation for the function $\delta(\theta)$:

\begin{equation}
\frac{\partial}{\partial {\theta}}[H(\theta)\sin\theta]= \frac{\cos\theta}{1-
\nu H(\theta)\cos\theta},
\label{main}
\end{equation}
where we have introduced the rescaled quantities:
\begin{equation}
H=\frac{\delta b\Delta}{2d_0}, \,\,\,\,  \nu=\frac{2Vd_0}{Db^2\Delta^3}.
\label{rescaled}
\end{equation}
We should find the smooth solution of this equation which starts from $H=0$ at 
$\theta=\theta_0$ and diverges as $H\rightarrow 1/(\nu\cos\theta)$ when  
$\theta\rightarrow \pi/2$. It turns out that this is an eigenvalue problem and 
such a solution exists only for specific values of the eigenvalue parameter $\nu$ 
which depends only on  $\theta_0$. Using Eqs. (\ref{triple},  \ref{rescaled}) 
we can present the final result for the steady-state growth velocity $V$ in the form:
\begin{equation}
V=\frac{Db^2\Delta^3}{2d_0}\nu(\Delta/2\varphi),
\label{result}
\end{equation}
where the scaling function $\nu(\Delta/2\varphi)$, obtained by the numerical solution 
of Eq. (\ref{main}),
 is presented in Fig. 3 and has the following asymptotic behavior:
\begin{equation}
\nu(\Delta/2\varphi)\approx 0.25,\,\,\,\,\,\, \Delta/\varphi\ll1
\label{small}
\end{equation}
and 
 \begin{equation}
\nu(\Delta/2\varphi)\approx 2.12(\Delta/2\varphi)^3,\,\,\,\,\,\, \Delta/\varphi\gg1
\label{large}
\end{equation}
Eqs. (\ref{result}), (\ref{small}) and (\ref{large}) are the main results of this 
paper
providing the expression for the steady-state velocity $V$ of the melting process 
along the grain boundary in terms of the driving force $\Delta$ and material 
parameters. 

\begin{figure}
\begin{center}
\epsfig{file=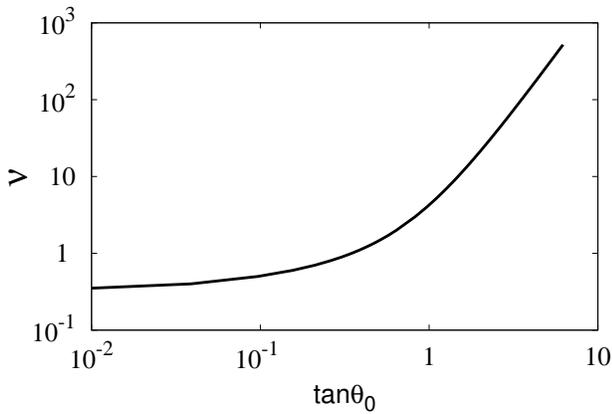, angle=-90,width=8cm}
\caption{The scaling function $\nu$ vs. $\tan\theta_0=\Delta/2\varphi$.}
\end{center}
\end{figure}

The analysis of this section is close in spirit to the analysis given in  
Ref. \cite{brener07acta} where we discussed the melting process in eutectic and 
peritectic systems. However, apart from very different physical ingredients 
involved,  the  lubrication approximations  used are quite different in these two 
cases.  In Ref. \cite{brener07acta} the assumption of small opening angles $\varphi$ 
immediately leads to  small variations of the angle $\theta$ in the vicinity of 
the value $\pi/2$. In the present analysis, the additional small parameter $\Delta$ 
leads to  wide variations of the angle $\theta$ depending on the ratio 
$\Delta/\varphi$, Eq. (\ref{triple}). Thus,  the lubrication approximation  developed 
here is more general than the approximation in Ref. \cite{brener07acta}. 
It involves calculations of the whole scaling function $\nu (\theta_0)$, while 
in \cite{brener07acta}  only a single eigenvalue parameter was calculated. 
Of course, in the limit $\theta_0\rightarrow\pi/2$ the present problem can be mapped
onto the problem of Ref. \cite{brener07acta}.

 \section{Kinetics of the grain boundary}
The kinetics of the grain boundary plays an important role
 allowing the necessary rotation of the structure in the vicinity of the triple junction 
 ( Fig. 2). The  angle $\theta$ along the grain boundary should change from its 
value $\theta_0$ at the triple junction to $\pi/2$ far ahead of the junction. 
Moreover, the concentration at the grain boundary should change from its value $c_S$ 
at the junction to the value $c_0$ far ahead.
The distribution of the concentration $c$ in the grain boundary is given 
by the solution of the  Chan equation, \cite{Chan}:
\begin{equation}
D_b\delta_b\partial^2{c}/\partial{s^2}-(c-c_0)V\cos\theta=0,
\end{equation}
where $D_b$ is the grain boundary diffusion coefficient,  $\delta_b$ is the thickness 
of the boundary. and $s$ is arc-length; $s=0$ at the triple junction 
and $s$ is negative along the grain boundary. It turns out that 
for small driving forces $\Delta$ the characteristic length 
of the concentration decay is much shorter than the characteristic length 
for the shape changes. This allows to use the  value $\theta_0$ instead of 
the current values $\theta (s)$.  Thus, we obtain:
$$c(s)-c_0 \approx (c_S-c_0)\exp(-k|s|),$$
\begin{equation}
k=\sqrt{V\cos\theta_0/D_b\delta_b}.
\label{distr}
\end{equation}
 
 The evolution of the grain boundary shape is controlled by   
 the recrystallization kinetics \cite{brener02acta} 
\begin{equation}
V\cos\theta=V_b[-d_bK_b+ b\Delta^2(s)],
\label{recr}
\end{equation}
where $V$ is the steady-sate velocity which we have already found from the analysis
of the melting process;
$V_b$ is the characteristic velocity scale proportional
 the grain boundary mobility, $d_b$ is the capillary length 
proportional to the surface tension of the boundary and 
$K_b=\partial\theta/\partial s$ is the curvature.  
The last term in Eq. (\ref{recr}) is due to the coherency strain energy effects and 
$\Delta(s)=[c(s)-c_0)]/(c_L-c_S)$, where $c(s)$ is given by the distribution 
Eq. (\ref{distr}). 
 Eq. (\ref{recr}) can be integrated once leading to:  
\begin{equation}
\theta(s)-\theta_0=-\frac{Vx(s)}{V_bd_b} +\frac{b\Delta^2}{2kd_b}[\exp(-2k|s|)-1],
\label{x}
\end{equation}
where $x(s)=\int_{0}^{s}ds\cos\theta <0$ (see Fig. 2b).
 Indeed, the characteristic length of the concentration decay scales as 
$k^{-1}\sim1/\sqrt{V}$ 
and it is much shorter than the characteristic length of the angle decay $\theta (s)$
 which scales as $1/V$ ( in the limit of small driving forces $\Delta$, 
 the steady-state velocity $V$ is also small according to Eq. (\ref{result})).
 We note also that the last term in Eq. (\ref{x}), 
which is proportional to $(D_b\Delta/D)^{1/2}$ ($kd_b\sim
(D/D_b)^{1/2}b\Delta^{3/2}$),  
is small in the limit of small driving forces $\Delta$. Thus, far ahead of the 
triple junction the position of the grain  boundary relative to the junction is given by 
(see Fig. 2b):  
$$x(-\infty)\approx -(\pi/2-\theta_0)V_bd_b/V.$$    
Moreover,  the diffusional flux along the 
grain boundary has a nonzero value
at the triple junction. In principal,  this flux should be taken into 
account  in the description of the diffusional field in the liquid phase. This effect 
has been neglected in the previous sections. More careful analysis shows that 
corrections to Eq. (\ref{V}) and (\ref{delta}) due to this effect are small 
in the limit  of small $\Delta\ll 1$.  
Thus, the  grain boundary 
kinetics plays an important role in necessary adjustment of this interface 
but has no influence on the melting kinetics as we have already discussed 
previously in Ref. \cite{brener07acta} in the context of the contact melting in 
eutectic and peritectic systems.

\section{Discussion}
\begin{figure}
\begin{center}
\epsfig{file=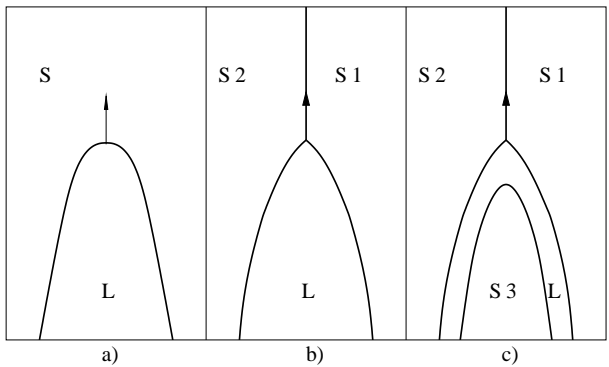, angle=0,width=8cm}
\caption{Schematic representation of the symmetric melting structures. 
a)- with only one melting front during the process in bulk of the grain; 
b)- with only one melting front during the process along the grain boundary; 
c)-  combined motion of the melting front along the grain boundary and the 
solidification front $S3/L$ via the LFM mechanism.  
}
\end{center}
\end{figure}

As we have already mentioned,  
 the process with only one melting front ( see Fig. 4a) would be controlled
 by the very slow diffusion in the mother solid phase. 
In this case the growth velocity
scales as in the classical dendritic growth. Assuming that the diffusion coefficient 
in the solid phase $D_s\ll D$, we can present this scaling in the form \cite{Langer}
\begin{equation} 
V\sim\frac{D_s^2\Delta^4}{Dd_0}\alpha^{7/4}.
\label{dendr}
\end{equation}
This scaling involves the additional small parameter $\alpha$ -
the anisotropy of surface tension.
For  melting along the grain boundary, Fig. 4b, 
one should ommit
the anisotropy factor \cite{brener07prl}: 
\begin{equation} 
V\sim\frac{D_s^2\Delta^4}{Dd_0}.
\end{equation}

 With the help of the LFM mechanism the system chooses  
a more efficient kinetic path to relax to 
the equilibrium state.
 For the case of two nearly parabolic fronts without grain boundary (see Fig.1), 
which was discussed in \cite{brener06prl}, the migration velocity scales as 
$$V\sim\frac{Db^2\Delta^3}{d_0}\alpha^{5/4}, \,\,\, \Delta\ll\alpha^{1/2},$$
\begin{equation}
V\sim\frac{Db^2\Delta^2}{d_0}\alpha^{7/4}, \,\,\, \Delta\gg\alpha^{1/2}.
\label{ref1}
\end{equation}

In the present paper we discuss  melting along the grain boundary via the LFM mechanism 
(Fig. 2b).   
The nucleation of the melt  takes place at the grain 
boundaries of the mother solid phase. Then the thin liquid layer extends along the 
grain boundary.
For small driving forces $\Delta$ and for not too small angles $\varphi$
the steady-state  velocity of this process scales, according to Eq. (\ref{result}), as 
\begin{equation} 
V\sim\frac{Db^2\Delta^3}{d_0}.
\label{mainscaling}
\end{equation}
If the nucleation of the melt takes place at the triple point where three grains meet 
together, the process could proceed as shown in Fig. 4c. This structure has not been 
investigated theoretically so far. However, we can guess that the velocity would scale 
as in Eq. (\ref{mainscaling}).  As we have already noted, the presence of the triple 
junction in this structure  produces a very strong 
perturbation of the solid-liquid interface and the anisotropy of the surface tension 
does not play an important role in such processes. Thus,  one can use the scaling 
given by the first relation in Eq. (\ref{ref1}), which describes the structure in Fig.1, 
formally setting $\alpha\sim 1$ \cite{brener07prl}. We note that the velocities of 
different processes increase from Eq. ({\ref{dendr}) to Eq. ({\ref{mainscaling}).    
 
Let us  estimate the velocity $V$ from Eq.(\ref{mainscaling}) and the thickness 
of the film $\delta\sim d_0/b\Delta$  from Eq.(\ref{rescaled})
using  characteristic values of the parameters: $D\sim 10^{-5}$cm$^2$/s, 
$d_0\sim 10^{-7}$cm, $b\sim 0.05$ and  $\Delta\sim 0.05$. 
This leads to $V\sim 10^{-4}$cm/s and $\delta\sim  10^{-4}-10^{-5}$cm. 

In conclusion, we have developed and analyzed a relatively simple model 
for the melting kinetics along the grain boundary in alloys. 
The process proceeds via LFM mechanism which is controlled by  relatively fast 
diffusion in the liquid film. 
The presence of the triple junction plays an important role in controlling 
the velocity of this process.
We solved this problem in lubrication approximation 
which allows to reduce the originally nonlocal problem to a local  and analytically 
tractable problem. We derived the expression for the  velocity  of
 the melting process along the grain boundary in terms of the driving force  
and material parameters.

We acknowledge the support by the Deutsche Forschungsgemeinschaft under 
project SPP 1126.


\begin{thebibliography}{99}
\bibitem{brener06prl} Brener EA, Temkin DE. Phys Rev Lett 2005;94:184501
\bibitem{yoon1} Yoon DN,   Hupmann WJ Acta Metall 1979;27: 973 
\bibitem{musch1} Muschik T,  Kaysser WA,  Hehenkamp T. 
Acta Metall 1989;37:603
\bibitem{yoon2} Yoon DN. Int. Mater. Rev 1995;40:149 
\bibitem{4}  Yoon DN,  Cahn JW,    Handwerker CA,  Blendell JE,
              Baik YJ. In: Interface Migration and Control of Microstructures.
                    Am Soc. Metals. Park. Ohio 1985, p. 19.
\bibitem{temkin2005}  Temkin DE.  Acta Mater 2005;53:2733
\bibitem{saito96}  Saito Y. Statistical Physics of Crystal Growth, World
                  Scientific Publishing, Singapore, 1996.
\bibitem{kessler88}  Kessler D,  Koplik J,  Levine H, 
             Adv Phys 1988;37:255
\bibitem{brener91}   Brener EA, and  Mel'nikov VI. Adv Phys 1991;40:53
\bibitem{temthes} Temkin DE. Abstracts of ICASP, June 7-10 2005, Stockholm, Sweden 
(KTH, Stockholm, 2005) 
\bibitem{brener07prl} Brener EA, Huter C, Pilipenko D, Temkin DE. Phys Rev Lett 
2007;99:105701 
\bibitem{brener07acta} Brener EA, Temkin DE. Acta Mater 2007;55:2785
\bibitem{Chan} Cahn JW. Acta Metall 1959;7:18
\bibitem{brener02acta} Brener EA, Temkin DE. Acta Mater 2002;50:1707
\bibitem{Langer} Barbieri A, Langer JS. Phys Rev A 1989;39:5314
\end{thebibliography}
\end{document}